\documentclass[journal=apchd5,manuscript=letter]{achemso}

\usepackage[version=3]{mhchem} 
\usepackage{textcomp,gensymb}
\usepackage{caption}
\usepackage{subcaption}
\usepackage{anyfontsize}
\usepackage{hyperref}

\author{Lifeng Chen}
\affiliation[University of Bristol]
{Department of Electrical and Electronic Engineering, University of Bristol, Merchant Venturers Building, Woodland Road, Bristol BS8 1UB, United Kingdom}
\email{lifeng.chen@bristol.ac.uk}

\author{Katrina A. Morgan}
\affiliation[University of Southampton]
{Optoelectronics Research Centre, University of Southampton, University Road, Southampton SO17 1BJ, United Kingdom}

\author{Ghada A. Alzaidy}
\affiliation[University of Southampton]
{Optoelectronics Research Centre, University of Southampton, University Road, Southampton SO17 1BJ, United Kingdom}

\author{Chung-Che Huang}
\affiliation[University of Southampton]
{Optoelectronics Research Centre, University of Southampton, University Road, Southampton SO17 1BJ, United Kingdom}
\email{cch@orc.soton.ac.uk}

\author{Ying-Lung Daniel Ho}
\affiliation[University of Bristol]
{Department of Electrical and Electronic Engineering, University of Bristol, Merchant Venturers Building, Woodland Road, Bristol BS8 1UB, United Kingdom}
\email{daniel.ho@bristol.ac.uk}

\author{Mike P. C. Taverne}
\affiliation[University of Bristol]
{Department of Electrical and Electronic Engineering, University of Bristol, Merchant Venturers Building, Woodland Road, Bristol BS8 1UB, United Kingdom}

\author{Xu Zheng}
\affiliation[University of Bristol]
{Department of Electrical and Electronic Engineering, University of Bristol, Merchant Venturers Building, Woodland Road, Bristol BS8 1UB, United Kingdom}

\author{Zhong Ren}
\affiliation[Oxford Instruments Plasma Technology]
{Oxford Instruments Plasma Technology, North End, Yatton, Bristol BS49 4AP, United Kingdom}

\author{Zhuo Feng}
\affiliation[University of Southampton]
{Optoelectronics Research Centre, University of Southampton, University Road, Southampton SO17 1BJ, United Kingdom}

\author{Ioannis Zeimpekis}
\affiliation[University of Southampton]
{Optoelectronics Research Centre, University of Southampton, University Road, Southampton SO17 1BJ, United Kingdom}

\author{Daniel W. Hewak}
\affiliation[University of Southampton]
{Optoelectronics Research Centre, University of Southampton, University Road, Southampton SO17 1BJ, United Kingdom}

\author{John G. Rarity}
\affiliation[University of Bristol]
{Department of Electrical and Electronic Engineering, University of Bristol, Merchant Venturers Building, Woodland Road, Bristol BS8 1UB, United Kingdom}

\title {Observation of Complete Photonic Bandgap in Low Refractive Index Contrast Inverse Rod-Connected Diamond Structured Chalcogenides}

\begin{document}

\begin{abstract}
Three-dimensional complete photonic bandgap materials or photonic crystals block light propagation in all directions.
The rod-connected diamond structure exhibits the largest photonic bandgap known to date and supports a complete bandgap for the lowest refractive index contrast ratio down to  $n_{high}/n_{low} \sim 1.9$.
We confirm this threshold by measuring a complete photonic bandgap in the infrared region in Sn--S--O $(n\sim1.9)$ and Ge--Sb--S--O $(n\sim2)$ inverse rod-connected diamond structures.
The structures were fabricated using a low-temperature chemical vapor deposition process via a single-inversion technique.
This provides a reliable fabrication technique of complete photonic bandgap materials and expands the library of backfilling materials, leading to a wide range of future photonic applications.
\end{abstract}
\bigskip{}
Three-dimensional (3D) complete photonic bandgap (PBG) structures have been widely studied since their invention in 1987 by John\cite{John1987} and Yablonovitch.\cite{Yablonovitch1987} A complete PBG structure can prohibit photon propagation in any direction and this strong confinement of light can be exploited for applications ranging through high precision sensing,\cite{Sunner2008} ultralow power and ultrafast optical switches,\cite{Nozaki2010} low threshold nanolasers,\cite{Khajavikhan2012} high efficiency single photon sources,\cite{Aharonovich2011} and integrated photonic circuits.\cite{John2012} However, such 3D PBG materials are difficult to fabricate. Currently two main techniques of fabrication have been demonstrated: bottom-up and top-down. The bottom-up method refers to schemes where nano-objects self-assemble into structures that then exhibit a PBG.\cite{Blanco2000,Galisteo-Lopez2011} The top-down approach refers to creating 3D structures using etching, ion-milling, lithography or laser writing that then produce PBGs.

Many of the top-down techniques involve miscellaneous fabrication steps such as wafer-fusion and micromanipulation,\cite{Aoki2009,Aoki2003} while others such as single prism holographic lithography\cite{Park2011,Campbell2000} do not allow for local modification for defects or waveguides.
Alternatively, direct laser writing (DLW) using two-photon polymerization (2PP) allows for a variety of high refractive index contrast (RIC) 3D photonic crystal (PhC) structures with complete PBGs in near-infrared\cite{Tetreault2006} and visible\cite{Frolich2013} regions to be realized.
To fulfill the high RIC $(> 2:1)$ requirement high index material (silicon or titanium dioxide) needs to be deposited into 3D templates. Most deposition temperatures are above the polymer melting point hence double inversion methods\cite{Varghese2013} or protective layers methods\cite{Frolich2013} have been used to make complete PBG materials.
There have also been successful demonstrations of DLW into photosensitive chalcogenide materials, followed by etching, showing bandgaps in the $3-4~\mu{}m$ wavelength range.\cite{Cumming2014,Nicoletti2011}

In this work, we have developed a low temperature Chemical Vapor Deposition (CVD) of chalcogenide materials\cite{Freeman2008,Hewak2010:Chalcogenide} to directly backfill unmodified polymer templates. Two materials were chosen for this work, Sn--S and Ge--Sb--S, due to their high refractive index values and low absorption in the near-infrared region, but also because they have attractive nonlinear optical properties suitable for applications, such as optical switches.\cite{Hewak2010:Chalcogenide} The chalcogenide materials are conformally coated on polymeric RCD templates,\cite{Chen2015} which are written by a commercial DLW system (Nanoscribe GmbH). The chalcogenide/polymer structure is then exposed to an oxygen plasma, resulting in selective etching of the polymeric scaffold. This novel approach results in chalcogenide inverse RCD structures, and here, we successfully demonstrate measurements showing a complete PBG at near-infrared wavelength $(0.9-1.7~\mu{}m)$ with low RIC ($n_{high}/n_{low} \sim 1.9:1$ for Sn--S--O/air and 2:1 for Ge--Sb--S--O/air) materials.

\begin{figure}
  \includegraphics[width=1\textwidth]{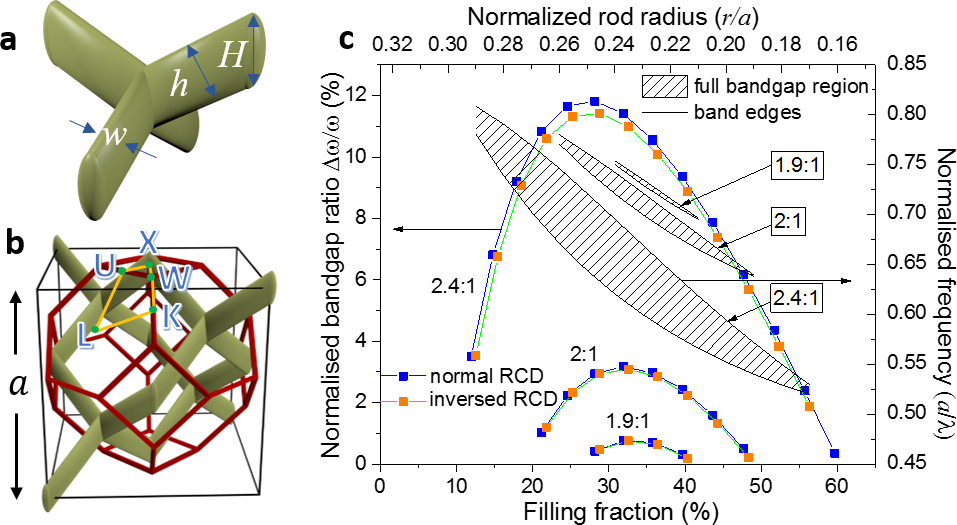}
  \caption{(a) Smallest unit of an RCD with structure parameters, width w and height h and H, labeled. (b) Individual RCD lattice with its Brillouin zone (red bold line), XWKLU are the symmetry points in reciprocal space.
  (c)
  Diagram calculated via MPB simulations: color dot lines showing complete bandgap ratio comparison of normal (blue) and inverse (orange) RCD in different high index material FF; bold black line with shadow in between indicates the complete PBG region changes as a function of normalized rod radius of RCD in an inverse RCD. The simulations assume rods are cylindrical and a high-index material with $n$ = 2.4, 2, and 1.9 in air ($n$ = 1), respectively.
  The rod radius was varied with a step of 0.01$a$.
  Note that the bandgap ratios were plotted as a function of high-index filling fraction (computed by MPB), while the bandgap regions were plotted as a function of the normalized radius $r/a$.
  The relationship between filling fraction and radius is almost, but not perfectly, linear.
  This explains the misalignment between the two types of plots.
  }
  \label{fig1}
\end{figure}

To achieve a complete PBG for smaller wavelengths, a higher PBG ratio (gap width to center wavelength ratio $\Delta \lambda / \lambda_{0}$) structure is required to minimize fabrication tolerance, that is, prevent errors from templates and depositions. The RCD structure,\cite{Chan1991} from the A7 crystal family,\cite{Chan1994} is reported to retain the highest complete PBG ratio among all crystal geometries and the lowest RIC ($n_{high}/n_{low} \sim 1.9:1$) known to support a complete PBG.\cite{Maldovan2004,Men2014:PhC-optimization} This RCD structure is described as rods replacing bonds between atoms in a diamond crystal. The conventional cubic unit cell consists of four tetrahedrons (Figure~\ref{fig1}a) stacked two by two in orthogonal directions (Figure~\ref{fig1}b). The lattice constant, $a$, is used to define an RCD structure (Figure~\ref{fig1}b). For a realistic simulation of the fabricated samples, elliptical rods are preferred, with a width $w$ and a "transverse height" $h$ ($\sim \sqrt{2/3} H$, where $H$ is the "vertical height"; Figure~\ref{fig1}a). The translational symmetry of an RCD (and its inversion) is the same as a face-centered cubic (FCC) structure, and the first Brillouin zone is a truncated octahedron (red line in Figure~\ref{fig1}b). XWKLU are the symmetry points on the Brillouin zone in an RCD structure. A high-quality RCD is not a layered structure and, thus, cannot be easily achieved using layer-by-layer 2D lithography methods. Moreover, the rod diameter required for direct high-index RCD\cite{Taverne2016} is below the 2PP DLW system resolution with a 780 $nm$ laser.
Fortunately, its inverse structure shows a slightly smaller PBG ratio of 11\% (compared to 11.7 \%) at the same RIC (2.4:1), material filling fraction (FF), and rod radius, as illustrated in Figure~\ref{fig1}c.
By utilizing this inverse structure, one can create a relatively low resolution (big rods) polymer template to realize an air-filled high index structure.

\section{Results and Discussion}

Using the MIT photonic band (MPB) software,\cite{Johnson2001:mpb} based on the plane wave expansion method, we calculated the normal and inverse RCD’s photonic band structure.
We also used Lumerical,\cite{website:Lumerical} a commercial-grade simulator based on the finite-difference time-domain (FDTD) method, to calculate their angular dependent reflection spectra.
For the FDTD simulations, we used a plane wave as source and set it to different propagation angles (relative to the normal incidence) to create angular spectrum results.
Substrates are not included in all calculations as the substrate thickness is far bigger than the photonic crystal thickness and it generates barely visible differences compared to the simulations without substrate, but hugely increases the required computational resources.
Figure~\ref{fig1}c plots the PBG ratio and the normalized frequency ($a/\lambda$) of the PBG position as a function of the air rod radius (or high index material FF) in an inverse RCD structure.
In this example, we assume rods are cylindrical and used three RICs 2.4:1, 2:1 and 1.9:1 (refractive index is averaged and nondispersive).
A band structure calculation for inverse RCDs with RIC 2.4:1 at different FF (Figure~\ref{fig1}c)  shows the complete PBG only appears within the range of air rod radius from 0.175$a$ to 0.3$a$, with a corresponding FF of material from 60\% to 10\% and normalized frequency of PBG from $0.525$ to $0.8(a/\lambda)$.
The maximum complete PBG ratio is 11\%, ranging from $0.63$ to $0.71(a/\lambda)$, and appears with a 0.25$a$ air rod radius.
When reducing the RIC to 2:1 and 1.9, the FF and radius ranges to obtain a complete PBG decrease, while the PBG ratio decreases and the midgap frequency increases. However, the optimal radius stays around 0.23-0.25$a$ for all three RICs.
A practical resolution constant for the RCD structure $a~=~1~\mu{}m$ is chosen, which results in the PBG wavelength range from $1.3$ to $1.9~\mu{}m$ and around 500~$nm$ diameter template rods.
An optimized commercialized DLW system based on 2PP has shown voxel resolution down to $200~nm$ lateral and $300~nm$ vertical.\cite{Hermatschweiler2007}

\begin{figure}
  \includegraphics[width=1\textwidth]{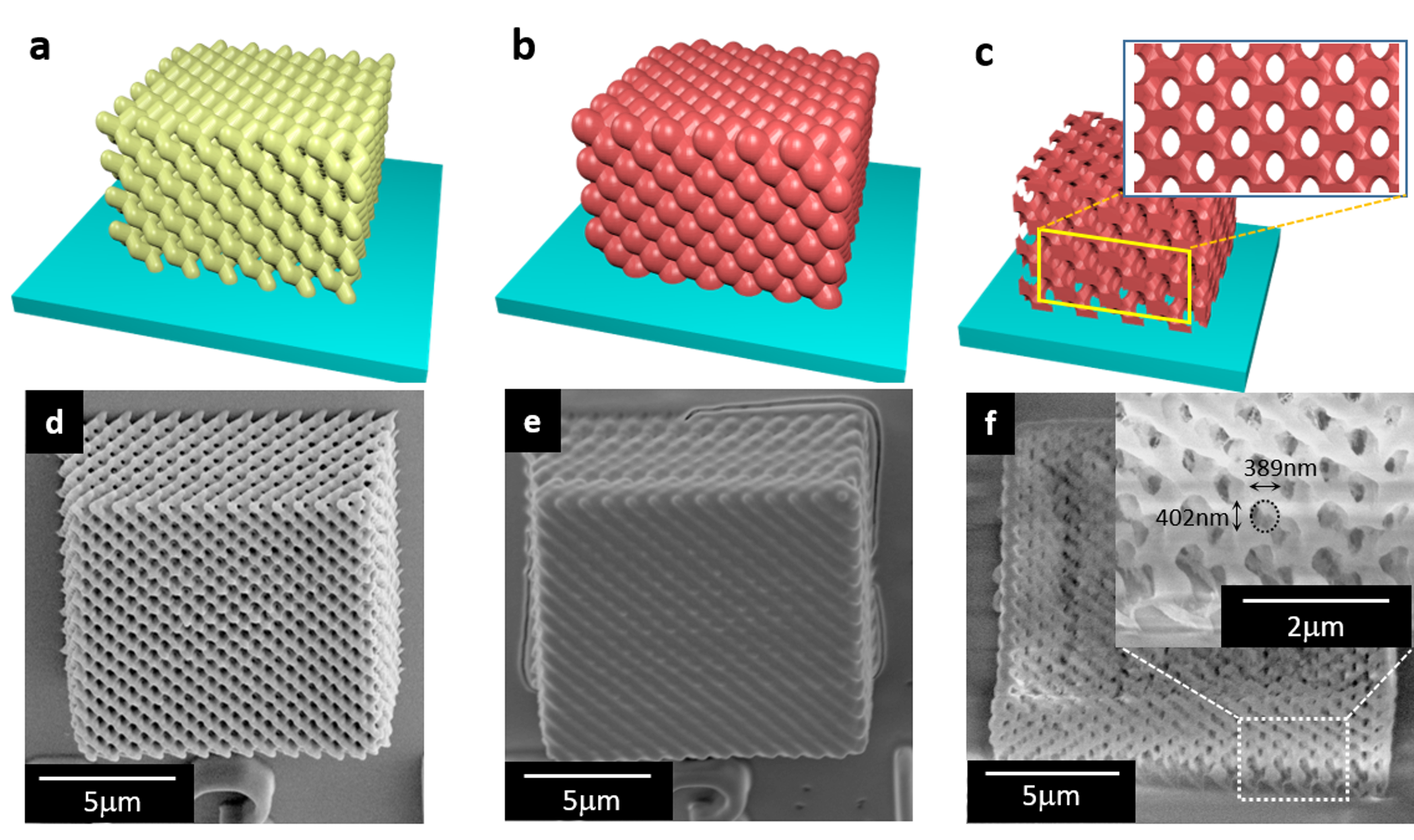}
  \caption{(a-c) Illustrations and (d-f) SEM photos for each fabrication step. (a) and (d) show the polymer template with a size around $14\times14~\mu{}m$ in the x-y plane. (b) and (e) backfilled (showing Ge--Sb--S structure) template with no visible air gap. (c) and (f) are the 45\degree{} oblique views for the inverse Ge--Sb--S--O RCD; insets show enlarged areas from the cross section, and parameters are measured as $w\sim389~nm$ and $H\sim402~nm$.}
  \label{fig2}
\end{figure}

The fabrication of an inverse RCD structure can be described in three main phases: template construction, high index material backfilling, and polymer removal, shown in Figure~\ref{fig2}. Phase one constructs a polymer RCD template (Figure~\ref{fig2}a) using a 2PP DLW method (see \nameref{sec:MaterialsAndMethods} for the details). We followed the optimized rod radius based on the MPB simulations (Figure~\ref{fig1}c). The fabricated template has 6 lattice periods in z-axis and 14 in x- and y-axes (Figure~\ref{fig2}d). To examine the 3D structure quality, an optical characterization of the template is performed prior to the backfilling process. This is done using a home-built Fourier imaging spectroscopy setup\cite{Chen2017} (see \nameref{sec:MaterialsAndMethods}).

In phase two (Figure~\ref{fig2}b), high refractive index chalcogenide materials (Sn--S or Ge--Sb--S) are conformally deposited into the polymer templates using an in-house built CVD system\cite{Huang2010} (see \nameref{sec:MaterialsAndMethods}). The deposition rate of the CVD materials is controlled by the ratios between precursors and reactive gas, chamber pressure, deposition temperature, and gas flow. The low melting point nature of the polymer template limits the deposition temperature to 200\celsius{} or below. The Sn--S deposition was carried out at room temperature whereas 150\celsius{} was used for Ge--Sb--S deposition. A Scanning Electron Microscope (SEM) photo of a fully backfilled Ge--Sb--S--polymer RCD structure is shown in Figure~\ref{fig2}e.

The final phase three is polymer removal. In phase two, the high index material grows omni-directionally inside and on top of the polymer templates. We used focused ion beam (FIB) milling to cut a thin layer off the top, to expose the buried polymer beneath, enabling the oxygen plasma to access the polymer template from above. The oxygen plasma reacts with the polymer to form gaseous compounds that escape from the template, but this treatment also partially oxidizes the chalcogenides resulting in reduced refractive indexes (see \nameref{sec:MaterialsAndMethods} for the details). Once the polymer template is removed, the partially oxidized high index materials will remain forming an inverse RCD structure. Figure~\ref{fig2}f shows the oblique view of a completed inverse RCD structure (Ge--Sb--S--O based) from SEM. Elliptical air gaps are visible in the cross section, with width $w\sim389~nm$ and height $h\sim501~nm$ (ascertained from the SEM measured $H\sim402~nm$ and Figure~\ref{fig1}a).

\begin{figure}
  \includegraphics[width=1\textwidth]{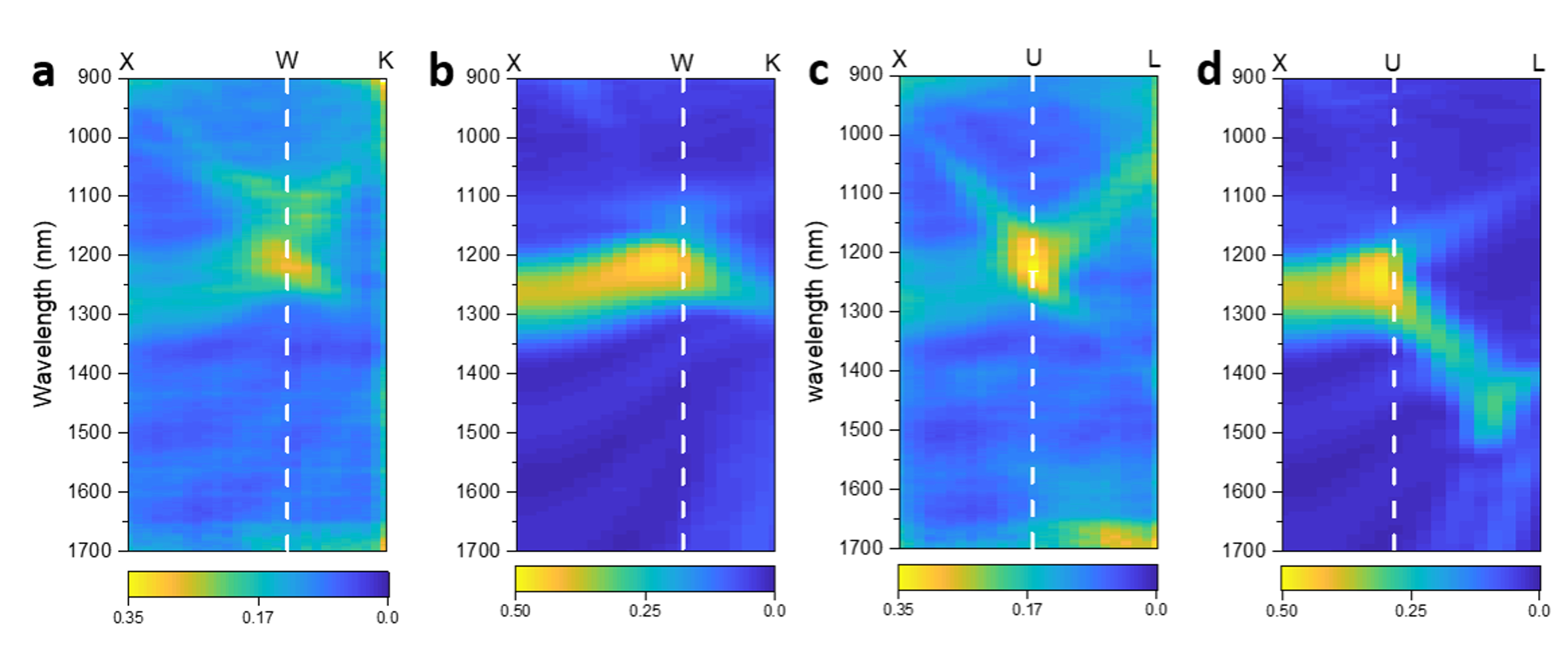}
  \caption{Intensity color plot for angular (a, c) reflection measurements and (b, d) FDTD simulation of RCD polymer templates. (a) and (b) are the results in X--W--K directions; (c) and (d) are the results in X--U--L directions.}
  \label{fig3}
\end{figure}

The photonic band structures of the polymer templates are measured using wide-angle Fourier imaging spectroscopy and compared with the FDTD simulations.\cite{Chen2017} The RCD template fabricated via the 2PP process is an air--polymer-based crystal, where the RIC is approximately 1:1.5. Figure~\ref{fig3} shows the angular reflection spectra comparison between measured polymer templates and simulations via FDTD.

Figure~\ref{fig3}a, b is the optical response (unpolarized reflection) for a detection angle in XUL and Figure~\ref{fig3}c,d is for a detection angle in XWK. Figure~\ref{fig3}a and c demonstrate partial band gaps with reflectivity above 20 \% (30 \% in simulation) at around $1200-1300~nm$ wavelength in normal incidence angle (X) and blue-shifted reflection peaks at the second symmetry point (U and W, respectively), with reflectivity up to about 40 \% in measurements and 50 \% in FDTD simulation. FDTD simulations use structure parameters $h~=~500~nm$, $w~=~400~nm$, $a~=~0.925~\mu{}m$ (slightly less than the targeted $1\mu{}m$ lattice size due to polymer shrinkage\cite{Chen2015}), and finite 10 by 10 lattice periods in the x--y plane and 6 periods in the z-direction. The simulated fundamental bandgaps closely match the experimental results from the polymer template, with minor differences appearing only in higher order bands, demonstrating a high-quality template has been achieved.

\begin{figure}[t]
  \includegraphics[width=1\textwidth]{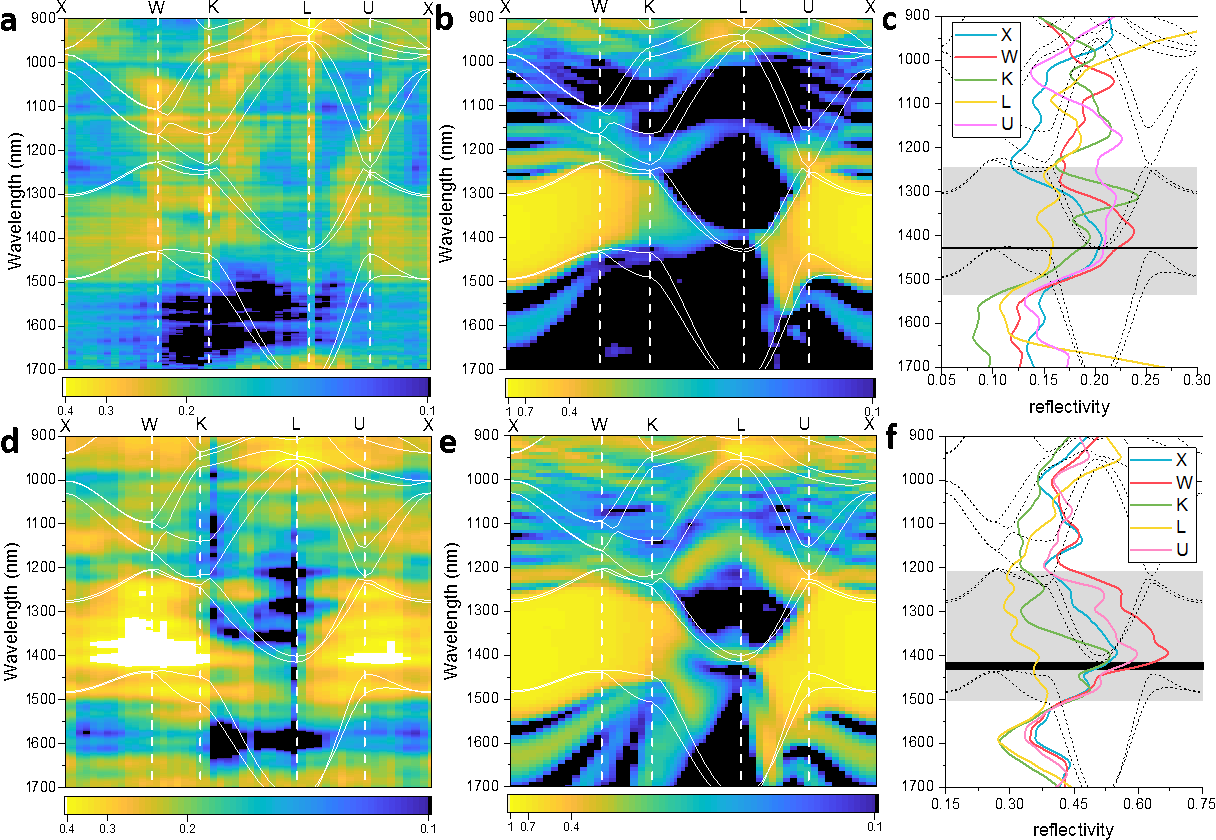}
  \caption{Color plot (a,~d) measured and (b,~e) FDTD simulated angular reflection spectra, mapped with photonic band diagram in X-W-K-L-U-X directions. The reflection spectra color line plots (c,~f) for measured structures of X, W, K, L, and U directions individually have been noise reduction processed to show the main features: the shadow area includes the main band features and the black area indicates the complete PBG; (a)--(c) are for Sn--S--O and (d)--(f) are for Ge--Sb--S--O.}
  \label{fig4}
\end{figure}

Figure~\ref{fig4}a, d, b, and e show the optical response of the Sn--S--O (Ge--Sb--S--O) inverse RCDs in all symmetry points XWKLUX for measurement and FDTD simulation, respectively.
MPB band structure calculations have been layered on top of all results in Figure~\ref{fig4}.
The refractive index of partially oxidized chalcogenides is estimated by using the energy-dispersive X-ray (EDX) spectroscopy technique to measure the material composition on both 3D structures and a test deposition on a wafer placed next to the sample in the deposition process.
The dispersions of index for both materials are less than 0.1 in the range of $900-1700~nm$, according to literature,\cite{Huang2010,El-Nahass2002} thus, we use averaged nondispersive refractive indexes for both materials in simulations. The resulting indexes are 1.9 for Sn--S--O and 2.0 for Ge--Sb--S--O at $900-1700~nm$ (see \nameref{sec:MaterialsAndMethods} for details).

In the measurement result for the Sn--S--O structure (Figure~\ref{fig4}a), a continuous reflection peak (the fundamental PBG) across symmetry points X--W--K and L--U--X appears at around $1250-1500~nm$ with $15 - 35\%$ reflectivity.
The fundamental band reflectivity at K--L direction drops to around 15\% due to the low RIC (1.9:1).
Some higher order bands (reflection peaks) appear between $1000-1300~nm$ in the W--K and U--X direction, matching its MPB simulation results.
For the Ge--Sb--S--O structure measurement result in Figure~\ref{fig4}d, a continuous reflection peak across all symmetry points appears at around $1250-1550~nm$, with a maximum 70~\% reflectivity at the W direction and lowest reflectivity around 30~\% at the L direction.
Both FDTD simulations (Figure~\ref{fig4}b, e) are based on finite size structures (10 lattices in the x--y plane).
A small DC offset was applied to the color scale of measured data to suppress the background scattered light.
The FDTD simulation parameters for the air rods are $h = 400~nm$ and $w = 450~nm$ for the Sn--S--O structure and $h = 500~nm$ and $w = 400~nm$ for Ge--Sb--S--O structure adjusted to best fit the optical results. This also confirms the values measured from the SEM results and supports our estimates of refractive indexes.
The FDTD simulation shows lowered reflections in the peak (drop to less than 20 \% reflectivity) at high observation angles (in the K--L region) due to the finite size (in the x--y plane) of the structures and edging effects, in contrast to the MPB (infinite structure size) calculations. For the same reasons, the measurements also show this effect, although there is a slight discontinuity seen around the L direction from a limitation of the imaging lens numerical aperture (NA). Figure~\ref{fig4}c and f demonstrate the reflection spectra at each symmetry point for Sn--S--O and Ge--Sb--S--O structures, respectively. The overlapping reflection peaks at around $1425~nm$ for the Sn--S--O structure, and $1410-430~nm$ for the Ge--Sb--S--O structure indicate the appearance of complete bandgaps in both Sn--S--O (bandgap ratio $> 0.3\%$) and Ge--Sb--S--O (bandgap ratio $> 1\%$).

\section{Conclusion}

By introducing the single-inversion process using a low-temperature CVD technique, we demonstrate low RIC (Sn--S--O and Ge--Sb--S--O) inverse RCD structures with complete PBGs working in the near-infrared region for the first time.
This single-inversion approach dramatically simplifies the 3D fabrication process.
Using low-temperature CVD and removing polymer templates with oxygen plasma, we have shown it is possible to completely fill the nanoscale void space of 3D templates with chalcogenide materials.
Optical modeling of the PBG material guided the design and, when compared with characterization results, enabled an estimation of device quality at each fabrication step. The complete PBGs ($> 0.3 \%$ and $> 1 \%$) of inverse RCDs formed in low RIC (1.9:1 and 2:1), via Sn--S--O and Ge--Sb--S--O chalcogenide materials, were experimentally measured with results compared with numerical simulations using the FDTD technique and plane-wave expansion method.
These results demonstrate the threshold of the lowest RIC supporting a complete PBG, experimentally validating results predicted by the topology optimization approach.\cite{Men2014:PhC-optimization} 

These results open the way for developing a process to reliably fabricate arbitrary photonic bandgap structures in technologically relevant wavelength regions ($1.4-1.6~\mu{}m$).
Moreover, our simulation work\cite{Taverne2016} points the way toward micro/nanocavity designs capable of confining light in mode volumes down to $10^{-3}$ cubic wavelengths.\cite{Taverne2018}
Our 3D lithography approach can be directly adapted to writing these cavity and waveguide structures, incorporating emitters to open new regimes of single photon level interactions,\cite{Kubo2010} novel nanolasers\cite{Khajavikhan2012} and high-bandwidth, lossless, and subwavelength scale optical circuits.\cite{John2012}

Future work will consider the fabrication of inverse RCD structures, along different growth directions (in particular, with the L direction aligned with the substrate normal) in order to measure the reflectivity at these higher angles more accurately.
However, the accuracy will then be affected by structural differences due to the elliptical voxel shape and the ensuing required differences in writing technique.

\section{Materials and Methods\label{sec:MaterialsAndMethods}}

\subsection{Direct Laser Writing (DLW)}

The DLW system is a commercial system from Nanoscribe GmbH, based on 2PP, which contains a $780~nm$ femtosecond laser (pulse width $\sim120~fs$ and repetition rate $\sim80~MHz$) and a high NA (=~1.4) oil immersion objective lens ($100\times$, Zeiss). The laser writing power is set to 20\% of mean output power (20~mW), with an adaptive piezo stage scanning speed of $50~\mu{}m/s$ and three repetitions per line. The photoresist used is a liquid negative resist, IP-L 780 (Nanoscribe GmbH), drop-casted onto a $22~mm\times22~mm\times170~\mu{}m$ glass substrate. An exposed template sample is developed using SU-8 developer for 30 min (to remove unpolymerized resist) and IPA for 5~min (to remove SU-8 developer).

\subsection{Fourier Imaging Spectroscopy (FIS)}

We used an identical system to that described in our previous work.\cite{Chen2017} This home-built Fourier imaging spectroscope uses a $4\times$ objective lens to collimate a fiber ($200~\mu{}m$ diameter) coupled white light source (Bentham Ltd. WLS100 $300-2500~nm$), focusing the light beam with an NA~=~0.9, $60\times$ objective lens on the sample. The detection plane is a projection image for the backfocal plane of the objective lens. This image is scanned by a fiber ($105~\mu{}m$ diameter) attaches to a x--y motorized stage, the other end of the fiber connects to a spectrometer (Ocean optics NIRQuest512), which has $900-1700~nm$ spectrum range. The angular resolution of the system is $\sim2\degree{}$ per scan step.

\subsection{Chemical Vapor Deposition (CVD)}

For Sn--S deposition:  $SnCl_{4}$ (99.999\% pure from Alfa Aesar) is used as the precursor to react with $H_{2}S$ gas (99.9\% pure from Air Liquide) to form Sn--S at room temperature with the chamber pressure of 100 mbar controlled by a Vacuubrand MV10NT diaphragm pump.  A 30~mm~O.D.~$\times{}$~1000~mm long quartz tube is used for CVD reaction and the precursor, $SnCl_{4}$ vapor, was delivered with $Ar$ gas through a mass flow controller (MFC) at 10~sccm, whereas $H_{2}S$ gas was delivered through another MFC at 50~sccm.

For Ge--Sb--S deposition: $GeCl_{4}$ (99.9999\% pure from Umicore) and $SbCl_{5}$ (99.999\% pure from GWI) are used as the precursors to react with $H_{2}S$ gas (99.9\% pure from Air Liquide) to form Ge--Sb--S at 150\celsius{} with atmospheric chamber pressure. A 30~mm~O.D.~$\times{}$~1000~mm long quartz tube is used for CVD reaction and the precursors, $GeCl_{4}$ and $SbCl_{5}$ vapors, are delivered individually with $Ar$ gas through MFCs at 20 and 80 sccm, respectively, whereas $H_{2}S$ gas is delivered through another MFC at 50 sccm.

\subsection{Inductively Coupled Plasma (ICP) Etching}

We used an ICP system, PlasmalabSystem 100 (ICP 180), from Oxford Instruments in the polymer removal process. The process was run twice for Sn--S structure and each time uses SEM to confirm complete removal of polymer template. The first run duration time was 5~min~30~s with 30 mTorr chamber pressure, oxygen flow rate was 50 sccm, and 100 W RF forward power and 400 W ICP forward power, reaction temperature was 60\celsius{}. The second run reduced duration to 5 min and chamber pressure to 20~mTorr, other settings unchanged. For Ge--Sb--S structure the ICP parameters changed to duration time 40~min, 30~mTorr chamber pressure, oxygen flow rate was 50~sccm, 20~W RF forward power and 400~W ICP forward power, reaction temperature decreased to 40\celsius{} to reduce etching rate.

\subsection{Evaluation of the Refractive Index Values}

\begin{figure}
  \includegraphics[width=0.43\textwidth]{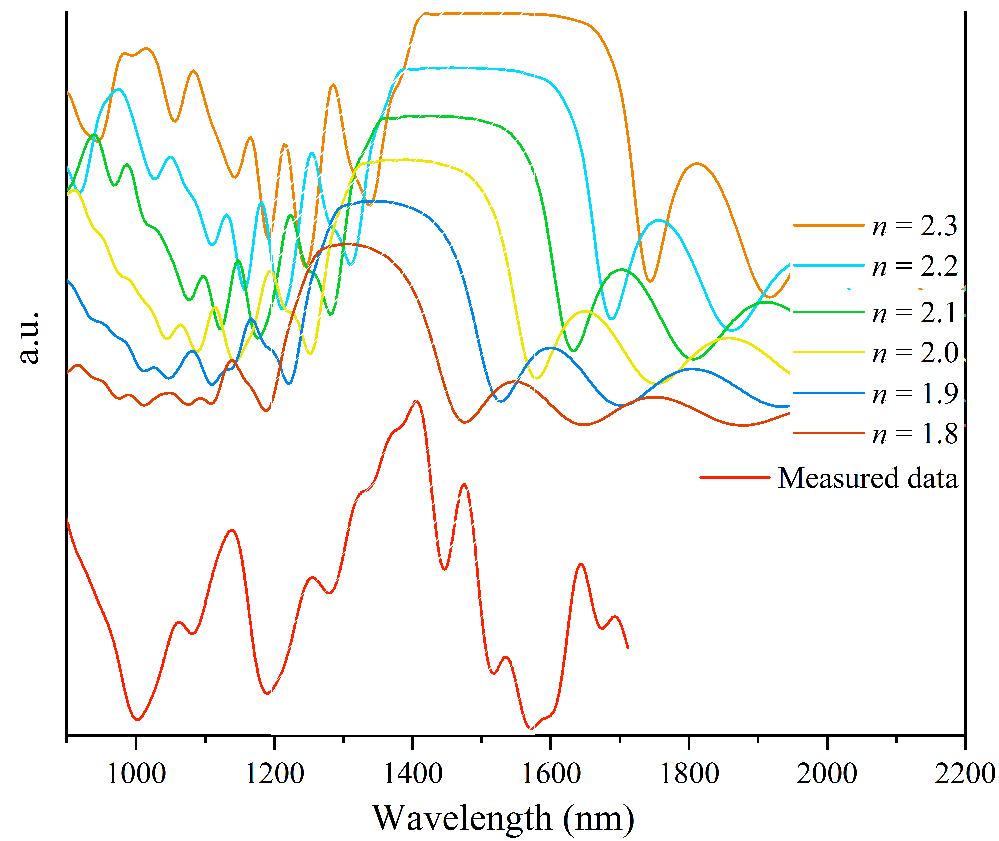}
  \caption{Ge--Sb--S--O refractive index fitted via FDTD simulations. The modeling results are compared with the measurement data (from the inverse RCD) at normal incident angle. The shadow region is where the PBG located. In this case, the refractive index of the structure is close to 2.0.}
  \label{fig5}
\end{figure}

\begin{figure}
  \includegraphics[width=0.43\textwidth]{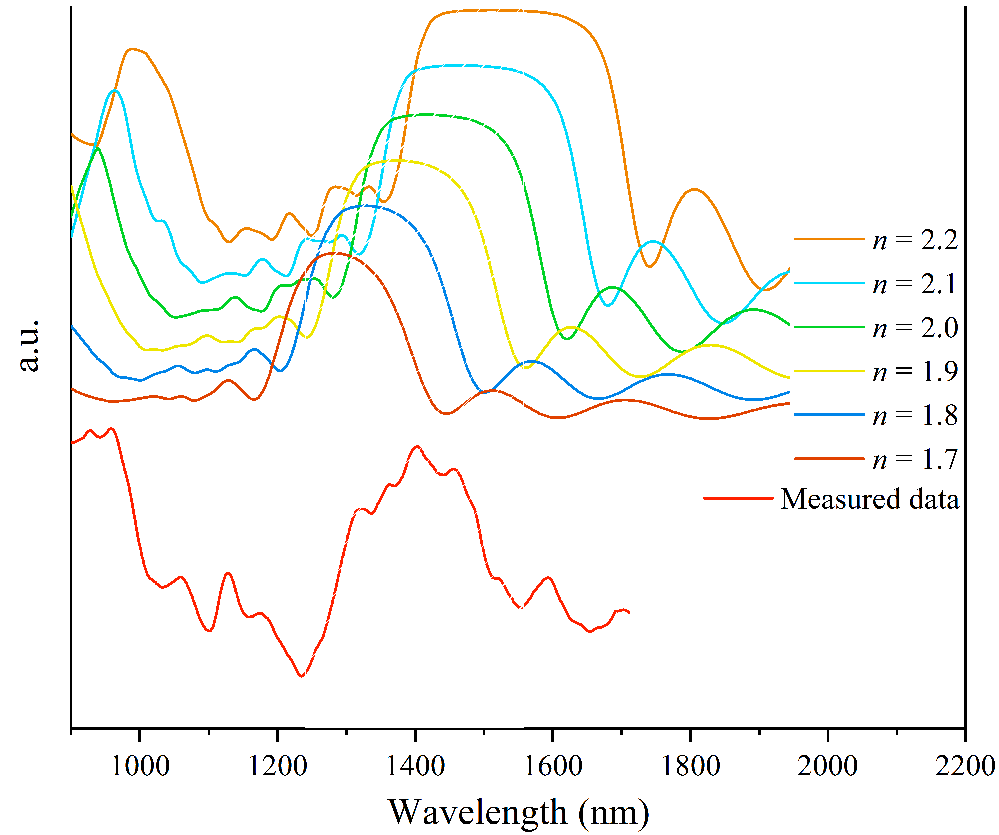}
  \caption{Sn--S--O refractive index fitted via FDTD simulations. The modeling results are compared with the measurement data (from the inverse RCD) at normal incident angle. The shadow region is where the PBG is located. In this case, the refractive index of the structure is close to 1.9.}
  \label{fig6}
\end{figure}

The refractive index of planar chalcogenide films grown under identical conditions and exposed to similar plasma etching were evaluated by ellipsometry.
However, these measurements yielded refractive index values that were far too high ($>2.6$ for GeSbS) to explain the optical measurements.
It was suspected that this might be due to oxidation effects which would be confined to the surface in planar films, while our porous structures are effectively fully oxidized.
EDX measurements of the composition of planar films confirmed limited oxidation.
In contrast, EDX measured compositions of 3D RCD structures were $Ge_{12}Sb_{15}S_{33}O_{40}$ and $Sn_{15}S_{14}O_{71}$, showing significant oxygen uptake and suggesting much lower refractive index values around 2.0 and 1.9, respectively.
This was further confirmed by using the refractive index (RI) values as a hand fitting parameter in the calculation of the expected reflection spectra at normal incidence using the FDTD method, as illustrated in Figures~\ref{fig5} and \ref{fig6}.

\clearpage{}

\section{Funding}

Engineering and Physical Sciences Research Council (EPSRC) (EP/M009033/1, EP/M008487/1, EP/M024458/1, EP/N00762X/1).

\begin{acknowledgement}

This work was carried out using the cleanroom fabrication facilities of the Centre for Nanoscience and Quantum Information (NSQI), University of Bristol, and the Optoelectronics Research Centre (ORC), University of Southampton, and computational facilities of the Advanced Computing Research Centre (ACRC), University of Bristol.

\end{acknowledgement}

\bibliography{main.bib}

%
%
%
%
%
%
%
%

\end{document}